# LogBase: A Scalable Log-structured Database System in the Cloud

Hoang Tam Vo [#1], Sheng Wang [#2], Divyakant Agrawal [†3], Gang Chen [§4], Beng Chin Ooi [#5]

[#]*National University of Singapore,* [†]*University of California, Santa Barbara,* [§]*Zhejiang University*

[1,2,5]{voht,wangsh,ooibc}@comp.nus.edu.sg, [3]agrawal@cs.ucsb.edu, [4]cg@cs.zju.edu.cn

## ABSTRACT

Numerous applications such as financial transactions (e.g., stock trading) are write-heavy in nature. The shift from reads to writes in web applications has also been accelerating in recent years. Write-ahead-logging is a common approach for providing recovery capability while improving performance in most storage systems. However, the separation of log and application data incurs write overheads observed in write-heavy environments and hence adversely affects the write throughput and recovery time in the system.

In this paper, we introduce LogBase – a scalable log-structured database system that adopts log-only storage for removing the write bottleneck and supporting fast system recovery. It is designed to be dynamically deployed on commodity clusters to take advantage of elastic scaling property of cloud environments. LogBase provides in-memory multiversion indexes for supporting efficient access to data maintained in the log. LogBase also supports transactions that bundle read and write operations spanning across multiple records. We implemented the proposed system and compared it with HBase and a disk-based log-structured record-oriented system modeled after RAMCloud. The experimental results show that LogBase is able to provide sustained write throughput, efficient data access out of the cache, and effective system recovery.

## 1. INTRODUCTION

There are several applications that motivate the design and implementation of LogBase, such as logging user activity (e.g., visit click or ad click from high volume web sites) and financial transactions (e.g., stock trading). The desiderata for the backend storage systems used in such write-heavy applications include:

- **High write throughput.** In these applications, a large number of events occur in a short period of time and need to be durably stored into the backend storage quickliest possible so that the system can handle a high rate of incoming data.

- **Dynamic scalability.** It is desirable that the storage systems are able to support dynamic scalability for the increasing workload, i.e., the ability to scale out and scale back on demand based on load characteristics.

- **Efficient multiversion data access.** The support of multiversion data access is useful since in these applications users often perform analytical queries on the historical data, e.g., finding the trend of stock trading or users' behaviors.

- **Transactional semantics.** In order to relieve application developers from the burden of handling inconsistent data, it is necessary for the storage system to support transactional semantics for bundled read and write operations that possibly access multiple data items within the transaction boundary.

- **Fast recovery from machine failures.** In large-scale systems, machine failures are not uncommon, and therefore it is important that the system is able to recover data and bring the machines back to usable state with minimal delays.

Storage systems for photos, blogs, and social networking communications in Web 2.0 applications also represent well-suited domains for LogBase. The shift from reads to writes has been accelerating in recent years as observed at Yahoo! [25]. Further, since such data are often written once, read often, and rarely modified, it is desirable that the storage system is optimized for high aggregate write throughput, low read response time, faut-tolerance and cost-effectiveness, i.e., less expensive than previous designs in storage usage while offering similar data recovery capability.

Previous designs for supporting data durability and improving system performance, which we shall discuss in more depth in Section 2, do not totally fit the aforementioned requirements. Copy-on-write strategy used in System R [14] incurs much overhead of copying and updating data pages, and therefore affects the write throughput. In POSTGRES [26], a delta record is added for each update, which would increase read latency since records have to be reconstructed from the delta chains. In write-ahead-logging (WAL) [19], in order to improve system performance while ensuring data durability, updates are first recorded into the log presumably stored in "stable storage", before being buffered into the memory, which can be flushed into data structures on disks at later time. We refer to this strategy as WAL+Data approach. Although this approach can defer writing data to disks, all the data have to be persisted into the physical storage eventually, which would result in the write bottleneck observed in write-heavy applications. In addition, the need to replay log records and update corresponding data structures when recovering from machine failures before the system becomes ready for serving new requests is another source of delay.

LogBase instead adopts log-only approach, in which the log serves as the unique data repository in the system, in order to remove the write bottleneck. The essence of the idea is that all write operations are appended at the end of the log file without the need of being reflected, i.e., updated in-place, into any data file. There are some immediate advantages from this simple design choice.

First, the number of disk I/Os will be reduced since the data only need to be written once into the log file, instead of being written into both log and data files like the WAL+Data approach. Second, all data will be written to disk, i.e., the log file, with sequential I/Os, which is much less expensive than random I/Os when performing in-place updates in data files. As a consequence, the cost of write operations with log-only approach is reduced considerably, and therefore `LogBase` can provide the much needed high write throughput for write-heavy applications. Log-only approach also enables cost-effective storage usage since the system does not need to store two copies of data in both log and data files.

Given the large application and data size, it is desirable that the system can be dynamically deployed in a cluster environment so that it is capable of adapting to changes in the workload while leveraging commodity hardware. `LogBase` adopts an architecture similar to HBase [1] and BigTable [8] where a machine in the system, referred to as tablet server, is responsible for some tablets, i.e., partitions of a table. However, `LogBase` is different in that it leverages the log as its unique data repository. Specifically, each tablet server uses a single log instance to record the data of the tablets it maintains. `LogBase` stores the log in an underlying distributed file system (DFS) that replicates data blocks across nodes in the cluster to guarantee that the probability of data loss is extremely unlikely, except catastrophic failures of the whole cluster. Consequently, `LogBase`'s capability of recovering data from machine failures is similar to traditional WAL+Data approach.

Since data, which are sequentially written into the log, are not well-clustered, it is challenging to process read operations efficiently. To solve this problem, tablet servers in `LogBase` build an index per tablet for retrieving the data from the log. Each index entry is a $<key, ptr>$ pair where $key$ is the primary key of the record and $ptr$ is the offset that points to the location of that record in the log. The index of each tablet can be maintained in memory since the size of an index entry is much smaller than the record's size. The in-memory index is especially useful for handling long tail requests, i.e., queries that access data not available in the cache, as it reduces I/O cost of reading index blocks. The interference of reads and writes over the log is affordable since reads do not occur frequently in write-heavy applications. As machines in commodity environments are commonly not equipped with dedicated disks for logging purpose, most scalable cloud storage systems such as HBase [1] also store both log and application data in a shared DFS and hence observe similar interferences.

`LogBase` utilizes the log records to provide multiversion data access since all data are written into the log together with their version number, which is the commit timestamp of the transactions that write the data. To facilitate reads over multiversion data, the indexes are also multiversioned, i.e., the $key$ of index entries now is composed of two parts: the primary key of the record as the prefix and the commit timestamp as the suffix. Furthermore, `LogBase` supports the ability to bundle a collection of read and write operations spanning across multiple records within transaction boundary, which is an important feature that is missing from most of cloud storage systems [7].

In summary, the contributions of the paper are as follows.

- We propose `LogBase` – a scalable log-structured database system that can be dynamically deployed in the cloud. It provides similar recovery capability to traditional write-ahead-logging approach while offering highly sustained throughput for write-heavy applications.

- We design a multiversion index strategy in `LogBase` to provide efficient access to the multiversion data maintained in the log. The in-memory index can efficiently support long tail requests that access data not available in the cache.

- We further enhance `LogBase` to support transactional semantics for read-modify-write operations and provide snapshot isolation – a widely accepted correctness criterion.

- We conducted an extensive performance study on LogBase and used HBase [1] and LRS, a log-structured record-oriented system that is modeled after RAMCloud [22] but stores data on disks, as our baselines. The results confirm its efficiency and scalability in terms of write and read performance, as well as effective recovery time in the system.

The paper proceeds as follows. In Section 2, we review background and related work. In Section 3, we present the design and implementation of `LogBase`. We evaluate the performance of `LogBase` in Section 4 and conclude the paper in Section 5.

## 2. BACKGROUND AND RELATED WORK

In this section, we review previous design choices for supporting data durability while improving system performance. We also discuss why they do not totally fit the aforementioned requirements of write-heavy applications.

### 2.1 No-overwrite Strategies

Early database systems such as System R [14] use shadow paging strategy to avoid the cost of in-place updates. When a transaction updates a data page, it makes a copy, i.e., a shadow, of that page and operates on that. When the transaction commits, the system records the changes to new addresses of the modified data pages. Although this approach does not require logging, the overheads of page copying and updating are much higher for each transaction, and adversely affect the overall system performance.

Another no-overwrite strategy for updating records is employed in POSTGRES [26]. Instead of performing in-place updates to the page, a delta record is added to store the changes from the previous version of the record. To perform a record reading request, the system has to traverse the whole chain from the first version to reconstruct the record, which affects the read performance considerably. Further, POSTGRES uses a force buffer policy, which requires the system to write all pages modified by a transaction into disk at commit time. Such high cost of write operations is inadequate for write-heavy applications.

### 2.2 WAL+Data

ARIES [19] is an algorithm designed for database recovery and enables no-force, steal buffer management, and thus improves system performance since updates can be buffered in memory without incurring "update loss" issues. The main principle of ARIES is write-ahead-logging (WAL), i.e., any change to a record is first stored in the log which must be persisted to "stable storage" before being reflected into the data structure.

WAL is a common approach in most storage systems ranging from traditional DBMSes, including open source databases like MySQL and commercial databases like DB2, to the emerging cloud storage systems such as BigTable [8], HBase [1] and Cassandra [16]. The main reason why this approach is popular is that while the log cannot be re-ordered, the data can be sorted in any order to exploit data locality for better I/O performance (e.g., data access via clustered indexes). However, this feature is not necessary for all applications, and the separation of log and application data incurs potential overheads that would reduce the write throughput and increase the time for system recovery.

In particular, although this design defers writing the application data to disks in order to guarantee system response time, all the data buffered in memory have to be persisted into the physical storage eventually. Therefore, the system might not be able to provide high write throughput for handling a large amount of incoming data in write-heavy applications. In addition, when recovering from machine failures the system needs to replay relevant log records and update corresponding data before it is ready for serving new user requests. As a consequence, the time for the system to recover from machine failures is delayed.

## 2.3 Log-structured Systems

Log-structured file systems (LFS) pioneered by Ousterhout and Rosenblum [24] for write-heavy environments have been well studied in the OS community. More recently, BlueSky [30], a network file system that adopts log-structured design and stores data persistently in a cloud storage provider, has been proposed.

Although `LogBase` employs the ideas of LFS, it provides a database abstraction on top of the segmented log, i.e., fine-grained access to data records instead of data blocks as in LFS. `LogBase` uses files, which are append-only, to implement its log segments, while LFS uses fixed size disk segments for its log. More importantly, `LogBase` maintains in-memory indexes for efficient record retrieval, and hence its log management is simpler than LFS as the log does not need to store metadata (e.g., inode structure) to enable random data access. To further facilitate database applications, `LogBase` clusters related records of a table during its log compaction for efficient support of clustering access.

Contemporary log-structured systems for database applications include Berkeley DB (Java Edition) and PrimeBase[1] – an open source log-structured storage engine for MySQL. Both systems are currently developed for single machine environment and use disk-resident indexes, which restricts system scalability and performance. Recent research systems for scalable log-structured data management include Hyder [5] and RAMCloud [22]. Hyder takes advantage of new advent of modern hardware such as solid-state drives to scale databases in a shared-flash environment without data partitioning. In contrast, `LogBase` aims to exploit commodity hardware in a shared-nothing cluster environment. Similarly, RAMCloud, which is a scalable DRAM-based storage system, requires servers with large memory and very high-speed network to meet latency goals, whereas `LogBase` is a disk-based storage system that is inherently designed for large-scale commodity clusters.

Following no-overwrite strategies introduced by early database systems, log-structured merge tree (LSM-tree) [21], which is a hierarchy of indexes spanning across memory and disk, is proposed for maintaining write-intensive and real-time indexes at low I/O cost. The log-structured history data access method (LHAM) [20] is an extension of LSM-tree for hierarchical storage systems that store a large number of components of the LSM-tree on archival media. bLSM-tree [25], an optimization of LSM-tree that uses Bloom filters to improve read performance, has been recently proposed. LSM-tree and bLSM-tree complement our work and can be exploited to extend the index capability of `LogBase` when the memory of a tablet server is scarce. We shall investigate this option in our experiments.

It is also noteworthy that LSM-trees are designed with the assumption that external write ahead logs are available. Therefore, although some cloud storage systems, such as HBase [1] and Cassandra [16], have adopted LSM-trees for maintaining their data, instead of performing in-place updates as in traditional DBMSes, they have not totally removed potential write bottlenecks since the separation of log and application data still exists in these systems.

## 3. DESIGN AND IMPLEMENTATION

In this section, we present various issues of the design and implementation of `LogBase` including data model, partitioning strategy, log repository, multiversion index, basic data operations, transaction management, and system recovery method.

### 3.1 Data Model

Cloud storage systems, as surveyed in [7], represent a recent evolution in building infrastructure for maintaining large-scale data, which are typically extracted from Web 2.0 applications. Most systems such as Cassandra [16] and HBase [1] employ key-value model or its variants and make a trade-off between system scalability and functionality. Recently, some systems such as Megastore [3] adopt a variant of the abstracted tuples model of an RDBMS where the data model is represented by declarative schemas coupled with strongly typed attributes. Pnuts [9] is another large-scale distributed storage system that uses the tuple-oriented model.

Since `LogBase` aims to provide scalable storage service for database-centric applications in the cloud, its data model is also based on the widely-accepted relational data model where data are stored as tuples in relations, i.e., tables, and a tuple comprises of multiple attributes' values. However, `LogBase` further adapts this model to support column-oriented storage model in order to exploit the data locality property of queries that frequently access a subset of attributes in the table schema. This adaptation is accomplished by the partitioning strategy presented in the below section.

### 3.2 Data Partitioning

`LogBase` employs vertical partitioning to improve I/O performance by clustering columns of a table into column groups which comprise of columns that are frequently accessed together by a set of queries in the workload. Column groups are stored separately in different physical data partitions so that the system can exploit data locality when processing queries. Such vertical partitioning benefits queries that only access a subset of columns of the table, e.g., aggregate functions on some attributes, since it saves significant I/O cost compared to the approach that stores all columns in the schema into a single physical table.

This partitioning strategy is similar to data morphing technique [15] which also partitions the table schema into column groups. Nevertheless, the main difference is that data morphing aims at designing a CPU cache-efficient column layout while the partitioning strategy in `LogBase` focuses on exploiting data locality for minimizing I/O cost of a query workload. In particular, given a table schema with a set of columns, multiple ways of grouping these columns into different partitions are enumerated. The I/O cost of each assignment is computed based on the query workload trace and the best assignment is selected as the vertical partitions of the table schema. Since we have designed the vertical partitioning scheme based on the trace of query workload, tuple re-construction is only necessary in the worst case. Moreover, each column group still embeds the primary key of data records as one of its componential columns, and therefore to reconstruct the tuple, `LogBase` collects the data in all column groups using the primary key as selection predicate.

To facilitate parallel query processing while offering scale out capability, `LogBase` further splits the data in each column group into horizontal partitions, referred to as tablets. `LogBase` designs the horizontal partitioning scheme carefully in order to reduce the number of distributed transactions across machines. In large-scale applications, users commonly operate on their own data which form an entity group or a key group [3, 12, 28]. By cleverly designing the key of records, all data related to a user could have the same key prefix, e.g., the user's identity. As a consequence, data accessed by a transaction are usually clustered on a physical machine. In this case, executing transactions is not expensive since the costly two-phase commit can be avoided.

---

[1]http://sourceforge.net/projects/pbxt/

For scenarios where the application data cannot be naturally partitioned into entity groups, we can implement a group formation protocol that enables users to explicitly cluster data records into key groups [12]. Another alternative solution is workload-driven approach for data partitioning [11]. This approach models the transaction workload as a graph in which data records constitute vertices and transactions constitute edges. A graph partitioning algorithm is used to split the graph into sub partitions while reducing number of cross-partition transactions.

### 3.3 Architecture Overview

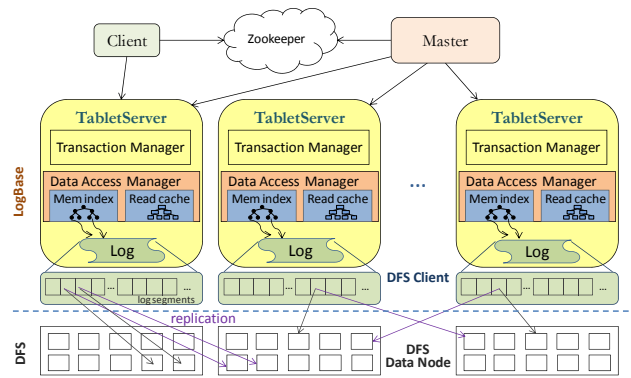

**Figure 1: System architecture.**

Figure 1 illustrates the overall architecture of LogBase. In this architecture, each machine – referred to as tablet server – is responsible to maintain several tablets, i.e., horizontal partitions of a table. The tablet server records the data, which might belong to the different tablets that it maintains, into its single log instance stored in the underlying distributed file system (DFS) shared by all servers. Overall, a tablet server in LogBase consists of three major functional layers, including transaction manager, data access manager, and log repository.

- **Log Repository.** At the bottom layer is the repository for maintaining log data. Instead of storing the log in local disks, the tablet servers employ a shared distributed file system (DFS) to store log files and provide fault-tolerance in case of machine failures. The implementation of Log Repository is described in Section 3.4.

- **Data Access Manager.** This middle layer is responsible to serve basic data operations including Insert, Delete, Update, and Get a specific data record. Data Access Manager also supports Scan operations for accessing records in batches, which is useful for analytical data processing such as programs run by Hadoop MapReduce[2]. In LogBase tablet severs employ in-memory multiversion indexes (cf. Section 3.5) for supporting efficient access to the data stored in the log. The processing of data operations is discussed in Section 3.6.

- **Transaction Manager.** This top layer provides interface for applications to access the data maintained in LogBase via transactions that bundles read and write operations on multiple records possibly located on different machines. The boundary of a transaction starts with a Begin command and ends with a Commit or Abort command. Details of transaction management is presented in Section 3.7.

The master node is responsible for monitoring the status of other tablet servers in the cluster, and provides the interface for users to update the metadata of the database such as create a new table and add column groups into a table. To avoid critical point of failures, multiple instances of master node can be run in the cluster and the active master is elected via Zookeeper [2], an efficient distributed coordination service. If the active master fails, one of the remaining masters will take over the master role. Note that the master node is not the bottleneck of the system since it does not lie on the general processing flow. Specifically, a new client first contacts the Zookeeper to retrieve the master node information. With that information it can query the master node to get the tablet server information and finally retrieve data from the tablet server that maintains the records of its interest. The information of both master node and tablet servers are cached for later user and hence only need to be looked up for the first time or when the cache is stale.

Although LogBase employs a similar architecture to HBase [1] and Bigtable [8], it introduces several major different designs. First, LogBase uses the log as data repository in order to remove the write bottleneck of the WAL+Data approach observed in write-heavy applications. Second, tablet servers in LogBase build an in-memory index for each column group in a tablet to support efficient data retrieval from the log. Finally, LogBase provides transactional semantics for bundled read and write operations accessing multiple records.

### 3.4 Log Repository

As discussed in Section 1, the approach that uses log as the unique data repository in the system benefits write-heavy applications in many ways, including high write throughput, fast system recovery and multiversion data access. Nevertheless, there could be questions about how this approach can guarantee the property of data durability in comparison to the traditional write-ahead-logging, i.e., WAL+Data approach.

GUARANTEE 1. *Stable storage. The log-only approach provides similar capability of recovering data from machine failures compared to the WAL+Data approach.*

Recall that in the WAL+Data approach, data durability is guaranteed with the "stable storage" assumption, i.e., the log file must be stored in a stable storage with zero probability of losing data. Unfortunately, implementing stable storage is theoretically impossible. Therefore, some methods such as RAID (Redundant Array of Independent Disks [23]) have been proposed and widely accepted to simulate stable storages. For example, a RAID-like erasure code is used to enable recovery from corrupted pages in the log repository of Hyder [5], which is a log-structured transactional record manager designed for shared flash.

To leverage commodity hardware and dynamic scalability designed for cluster environment, LogBase stores the log in HDFS[3] (Hadoop Distributed File System). HDFS employs $n$-way replication to provide data durability ($n$ is configurable and set to 3-way replication as default since it has been a consensus that maintaining three replicas is enough for providing high data availability in distributed environments). The log can be considered as an infinite sequential repository which contains contiguous segments. Each segment is implemented as a sequential file in HDFS whose size is also configurable. We set the default size of segments to 64 MB as in HBase [1].

---
[2]http://hadoop.apache.org/mapreduce

[3]http://hadoop.apache.org/hdfs

Replicas of a data block in HDFS are synchronously maintained. That is, a write operation to a file is consistently replicated to $n$ machines before returning to users. This is equivalent to RAID-1 level or mirroring disks [23]. Further, the replication strategy in HDFS is rack-aware, i.e., it distributes replicas of a data block across the racks in the cluster, and consequently guarantees that the probability of data loss is extremely unlikely, except catastrophic failures of the whole cluster. Therefore, the use of log-only approach in LogBase does not reduce the capability of recovering data from machine failures compared to the other systems. Note that HBase [1] also stores its log data (and its application data) in HDFS.

Each tablet server in LogBase maintains several tablets, i.e., partitions of a table, and record the log data of these tablets in HDFS. There are two design choices for the implementation of the log: (i) a single log instance per server that is used for all tablets maintained on that server and (ii) the tablet server maintains several log instances and each column group has one log instance. The advantages of the second approach include:

- **Data locality.** Since LogBase uses log as the unique data repository, it needs to access the log to retrieve the data. If a log instance contains only the data that are frequently access together, e.g., all rows of a column group, it's likely to improve the I/O performance for queries that only access that column group. On the contrary, in the first approach, the system needs to scan the entire log containing rows of all column groups.

- **Data recovery.** If a tablet server fails, its tablets will be assigned to other servers. In the second approach, one log represents one column group, and therefore, other servers only need to reload the corresponding index file and check the tail of that log (from the consistent point immediate after the latest checkpoint). Otherwise, in the first approach, the log has to be sorted and split by column group, and then scanned by the corresponding servers as in BigTable [8] and HBase [1].

However, the downside of the second approach is that, the underlying distributed file system has to handle many read/write connections that are used for multiple log instances. In addition, it also consumes more disk seeks to perform writes to different logs in the physically storage. Since LogBase aims at write-heavy applications that require sustained write throughput, we choose the first approach, i.e., each tablet server uses a single log instance for storing the data from multiple tablets that it maintains. Moreover, this approach still can support data locality after the log compaction process (cf. Section 3.6.5) which periodically scans the log, removes out-of-date data and sorts the log entries based on column group, primary key of the record, and timestamp of the write. That is, all data related to a specific column group will be clustered together after the log compaction.

A log record comprises of two components $<LogKey, Data>$. The first component, $LogKey$, stores the information of a write operation, which includes log sequence number (LSN), table name, and tablet information. LSN is used to keep track of updates to the system, and is useful for checkpointing and recovery process (cf. Section 3.8). LSN either starts at zero or at the last known LSN persisted in the previous consistent checkpoint block. The second component, $Data$, is a pair of $<RowKey, Value>$ where $RowKey$ represents the id of the record and $Value$ stores the content of the write operation. $RowKey$ is the concatenation of the record's primary key and the column group updated by the write operation, along with the timestamp of the write. Log records are to be persisted into the log repository before write operations can return to users.

### 3.5 In-memory Multiversion Index

Since LogBase records all writes sequentially in the log repository, there is no clustering property of data records stored on disks. As a result, access to data records based on their primary keys is inefficient as it is costly to scan the whole log repository only for retrieving some specific records. Therefore, LogBase builds indexes over the data in the log to provide efficient access to the data.

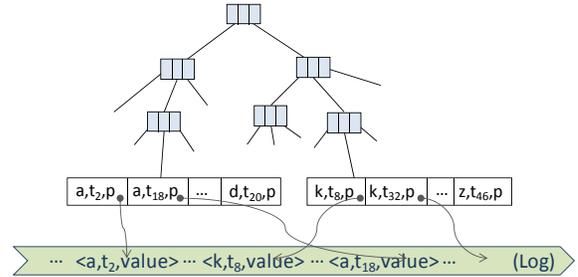

**Figure 2: Multiversion index over the log repository.**

In particular, tablet servers build a multiversion index, as illustrated in Figure 2, for each column group in a tablet. LogBase utilizes the log entries to provide multiversion data access since all data are written into the log together with their version numbers, i.e., the timestamp of the write. To facilitate reads over multiversion data, the indexes are also multiversioned. The indexes resemble $B^{link}$-trees [17] to provide efficient key range search and concurrency support. However, the content of index entries is adapted to support multiversion data. In our indexes, each index entry is a pair of $<IdxKey, Ptr>$. The $IdxKey$ is composed of two parts: the primary key of the record as the prefix and the timestamp as the suffix. $Ptr$ is the offset that points to the location of a data record in the log, which includes three information: the file number, the offset in the file, the record's size.

We design an index key as a composite value of record id and timestamp so that the search for current as well as historical versions of particular data records, which is the major access pattern in our applications, can be done efficiently. Historical index entries of a given record id, e.g., key $a$ in Figure 2, are clustered in the index and can be found by performing an index search with the data key $a$ as the prefix. Among the found entries, the one that has the latest timestamp contains the pointer to the current version of the data record in the log.

The ability to search for current and historical versions efficiently is useful for developing the multiversion concurrency control in LogBase (cf. Section 3.7). Although multiversion indexes can be implemented with other general multiversion access methods, e.g., Time-Split B-tree (TSB-tree) [18], these methods are mainly optimized for temporal queries by partitioning the index along time and attribute value dimensions, which increases the storage space and insert cost considerably.

The indexes in LogBase can be stored in memory since they only contain the $<IdxKey, Ptr>$ pairs whose size are much smaller than the record's size. For example, while the size of records, e.g., blogs' content or social communications, could easily exceed 1 KB, the $IdxKey$ only consumes about 16 bytes (including the record id and timestamp of long data type) and $Ptr$ consumes about 8 bytes (including the file number and record size as short data type, and the file offset as integer data type), which makes a total size of 24 bytes each index entry. Assuming that the tablet server can reserve 40% of its 1 GB heap memory for in-memory indexes (HBase [1] uses a similar default setting for its memtables), the indexes of that server can maintain approximately 17 million entries.

There are several methods to scale out `LogBase`'s index capability. A straight-forward way is to increase either the heap memory for the tablet server process or the percentage of memory usage for indexes (or both). Another solution is to launch more tablet server processes on other physical machines to share the workload. Finally, `LogBase` can employ a similar method to log-structured merge-tree (LSM-tree) [21] for merging out part of the in-memory indexes into disks, which we shall investigate in the experiments.

A major advantage of the indexes in `LogBase` is the ability to efficiently process long tail requests, i.e., queries that access data not available in read cache. `LogBase` uses in-memory indexes for directly locating and retrieving data records from the log with only one disk seek, while in the WAL+Data approach (e.g., in HBase [1]) both application data and index blocks need to be fetched from disk-resident files, which incurs more disk I/Os.

The downside of in-memory indexes is that their content are totally lost when machines crash. To recover the indexes from machine failures, the restarted server just scans its log and reconstructs the in-memory index for the tablets it maintains. In order to reduce the cost of recovery, `LogBase` performs checkpoint operation at regular times. In general, tablet servers periodically flush the in-memory indexes into the underlying DFS for persistence. Consequently, at restart time the tablet server can reload the indexes quickly from the persisted index files back into memory. We describe the details of `LogBase`'s recovery technique in Section 3.8.

## 3.6 Tablet Serving

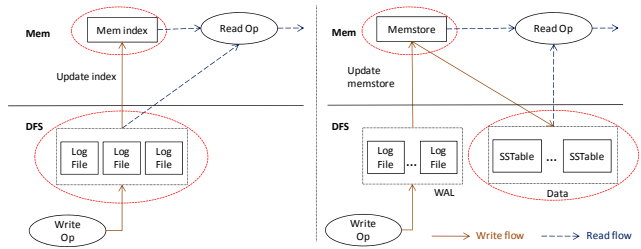

**Figure 3: Tablet serving of LogBase (left) vs. HBase (right).**

Now we present the details of a tablet server in `LogBase`, which uses only log files to facilitate both data access and recovery. As illustrated in Figure 3, each tablet server manages two major components, including (i) the single log instance (consisting of sequential log segments) which stores data of multiple tablets maintained by the server, and (ii) the memory index for each column group which map the primary key of data records to their location in the log. Another major component (not shown) is the transaction manager whose details will be described in the next section.

`LogBase` differs from HBase [1] on every aforementioned component. More specifically, HBase stores data in data files which are separate with the log and uses memtables to buffer recently updated data, in addition to the fact that it does not support transactional semantics for bundled read and write operations. The benefits of log-only approach compared to WAL+Data approach when serving write-heavy applications have been briefly discussed in Section 1. In the following, we shall describe how `LogBase` performs basic data operations such as write, read, delete, and scan over the tablets as well as tablet compaction operation.

### 3.6.1 Write

When a write request (`Insert` or `Update`) arrives, the request is first transformed into a log record of $< LogKey, Data >$ format, where $LogKey$ contains meta information of the write such as log sequence number, table name, and tablet information while $Data$ stores the content of the write, including the record's primary key, the updated column group, the timestamp of the write, and the new value of data. Then the tablet server writes this log record into the log repository.

After the log record has been persisted, its starting offset in the log along with the timestamp are returned so that the tablet server subsequently updates the in-memory index of the corresponding updated column group. This guarantees that the index are able to keep track of historical versions of the data records. The indexes are used to retrieve the data records in the log at later time.

In addition, the new version of data can also be cached in a read buffer (not shown in Figure 3) so that `LogBase` can efficiently serve read requests on recently updated data. While the in-memory index is a major component and is necessary for efficient data retrieval from the log, read buffer is only an optional component whose existence and size are configurable parameters. The read buffer in `LogBase` is different from the memtable in HBase [1] in that the read buffer is only for improving read performance while the memtable stores data and needs to be flushed into disks whenever the memtable is full, which incurs write bottlenecks in write-intensive applications.

A counter is maintained to record the number of updates that have occurred to the column group of a tablet. If the number of updates reaches a threshold, the index can be merged out into an index file stored in the underlying DFS and the counter is reset to zero. Persisting indexes into index files helps to provide a faster recovery from failures, since the tablet servers do not need to scan the entire log repository in order to rebuild the indexes. Note that the DFS with 3-way synchronous replication is sufficient to serve as a stable storage for index files (as the case of log files and discussed in Section 3.4).

### 3.6.2 Read

To process a `Get` request, which retrieves data of a specific record given its primary key, the tablet server first checks whether the corresponding record exists in the read buffer. If the value is found, it is returned and the request is completed. Otherwise, the server obtains the log offset of the requested record from the in-memory index. With this information, the data record is retrieved from the log repository, and finally returned to clients. By default, the system will return the latest version of the data of interest. To access historical versions of data, users can attach a timestamp $t_q$ with the retrieval request. In this case, `LogBase` fetches all index entries with the requested key as the prefix and follows the pointer of the index entry that has the latest timestamp before $t_q$ to retrieve the data from the log.

Meanwhile, the read buffer also caches the recent fetched record for serving possible future requests. Since there is only one read buffer per tablet server and the size of the read buffer is limited, an effective replacement strategy is needed to guarantee the read buffer is fully exploited while reducing the number of cache misses. In our implementation, we employ the LRU strategy which discards the least recently used records first. However, we also design the replacement strategy as an abstracted interface so that users can plug in new strategies that fit their application access patterns. With the use of read buffer, `LogBase` can quickly answer queries for data that have recently been updated or read, in addition to the ability to process long tail requests efficiently via in-memory indexes.

Note that the vertical partitioning scheme in `LogBase`, as discussed in Section 3.2, is designed based on the workload trace, and therefore most queries and updates will access data within a column group. In the case where tuple reconstruction is necessary, `LogBase` collects componential data of a record from all corresponding column groups.

### 3.6.3 Delete

A tablet server in `LogBase` performs a *Delete* operation given a record primary key in two steps. First, it remove all index entries associated with this record key from the in-memory index. By doing this all incoming queries at later time cannot find any pointer from the index in order to access the data record in the log repository. However, in the event of tablet server's restart after failures, the index is typically reloaded from the previous consistent checkpoint file, which still contains the index entries that we have attempted to remove in the first step.

Therefore, in order to guarantee durable effect of the *Delete* operation, `LogBase` performs a second step which persists a special log entry, referred to as invalidated log entry, into the log repository to record the information about this *Delete* operation. While this invalidated log entry also contains *LogKey* similar to normal log entries, its *Data* component is set to *null* value in order to represent the fact that the corresponding data record has been deleted. As a consequence, during the restart of the tablet server, this invalidated log entry will be scanned over and its deletion effect will be reflected into the in-memory index again.

### 3.6.4 Scan

`LogBase` supports two types of scan operations, including range scan and full table scan. A range scan request takes a start key and an end key as its input. If the query range spans across tablet servers, it will be divided into subranges which are executed in parallel on multiple servers. Each tablet server process a range scan as follows. First, it traverses the in-memory index to enumerate all index entries that satisfies the query range. Then, it follows the pointers in the qualified index entries to retrieve the data from the log repository. Since the data in the log are not clustered based on the search key, it is not efficient when handling with large range scan queries. However, `LogBase` periodically performs log compaction operation which will be discussed below. After this compaction, data in the log are typically sorted and clustered based on the data key. Therefore, `LogBase` can support efficient range scan queries, i.e., clustering access on the primary key of data records, if the log compaction operation is performed at regular times.

In contrast to range scan queries, full table scans can be performed efficiently in `LogBase` without much optimization. Since full table scans do not require any specific order of access to data records, multiple log segments, i.e., log files, in the log repository of tablet servers are scanned in parallel. For each scanned record, the system checks its stored version with the current version maintained in the in-memory index to determine whether the record contains latest data.

### 3.6.5 Compaction

In the log-only approach, updates (and even deletes) are sequentially appended as a new log entry at the end of the log repository. After a period of time, there could be obsolete versions of data that are not useful for any query, but they still consume storage capacity in the log repository. Therefore, it is important to perform a vacuuming process, referred to as compaction, in order to discard out-of-date data and uncommitted updates from the log repository and reclaim the storage resources.

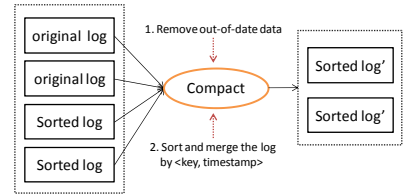

**Figure 4: Log compaction.**

Compaction could be done periodically as background process or more frequently when the system has spare CPU and I/O bandwidth. Figure 4 illustrates the compaction process performed by a tablet server in `LogBase`. In particular, `LogBase` performs a MapReduce-like job which takes the current log segments (some of them are sorted log segments, resulted from the previous compaction) as its input, removes all obsolete versions of data and invalidated records, and finally sorts the remaining data based on the following criteria (listed from the highest to lowest priority): table name, column group, record id, and timestamp. The result of this job is a set of sorted log segments in which data are well-clustered. Then, each tablet server builds the in-memory indexes over these new log segments. After the indexes have been built, the tablet server now can efficiently answer clients' queries on the clustered data in the sorted log segments.

Note that until this time point, old log segments and in-memory indexes are still in use and all clients' update requests from the start of the running compaction process are stored in new log segments which will be used as inputs in the next round of compaction. That is, `LogBase` can serve clients' queries and updates as per normal during the compaction process. After the compaction process has finished, i.e., the resulted sorted segments and in-memory indexes are ready, the old log segments and in-memory indexes can be safely discarded.

An additional optimization is adopted during the compaction process to decrease the storage consumption of log segments and further improve I/O performance for queries. Specifically, since the data in the resulting log segments are clustered by table name and column group already, it is not necessary to store this information in every log entries any more. Instead, the tablet server only needs to maintain a metadata which maps the table name and column group information to a list of log segments that store its data.

## 3.7 Transaction Management and Correctness Guarantees

In the previous section, we have presented `LogBase`'s basic data operations, which only guarantee single row ACID properties similar to other cloud storage systems such as Pnuts [9], Cassandra [16] and HBase [1]. We now present how `LogBase` ensures ACID semantics for bundled read and write operations spanning across multiple records.

### 3.7.1 Concurrency Control and Isolation

**The Rationale of MVOCC.** Recall that `LogBase` is designed with a built-in function of maintaining multiversion data. In addition, the careful design of the data partitioning scheme in `LogBase`, which is based on application semantics and query workload, clusters data related to a user together, and thus reduces the contention between transactions as well as the number of distributed transactions. Consequently, we employ a combination of multi-version and optimistic concurrency control (MVOCC) to implement isolation and consistency for transactions in `LogBase`.

A major advantage of MVOCC is the separation of read-only and update transactions so that they will not block each other. In particular, read-only transactions access a recent consistent snapshot of the database while update transactions perform on the latest version of the data. Therefore, read-only transactions always commit successfully, whereas an update transaction after finishing its read phase has to validate its possible conflicts with other concurrently executing update transactions before being allowed to enter the write phase.

While traditional OCC needs to maintain old write-sets of committed transactions in order to verify data conflicts, the MVOCC in `LogBase` provides another advantage that in the validation phase of update transactions, the transaction manager can use the version numbers of data records to check for conflicts with other update transactions. In particular, to commit an update transaction $T$, the transaction manager checks whether $T$'s write set are updated by other concurrent transactions that have just committed by comparing the versions of the records in $T$'s write set that $T$ has read before (there is no blind write) with the current version of the records maintained in the in-memory indexes. If there is any change in the record versions, then the validation fails and $T$ is restarted. Otherwise, the validation return success and T is allowed to enter the write phase and commit.

**Validation with Write Locks.** To avoid possible conflicts of concurrent writes, `LogBase` embeds write locks into the validation phase of MVOCC. In particular, an update transaction first executes its read phase as per normal; however, at the beginning of validation phase, the transaction manager will request write locks over the data records for its intention writes. If all the locks can be obtained and the validation succeeds, the transaction can execute its write phase, and finally release the locks. Otherwise, if the transaction manager fails to acquire all necessary write locks, it will still hold the existing locks while re-executing the read phase and trying to request again the locks that it could not get in the first time. This means that the transaction keeps pre-claiming the locks until it obtains all the necessary locks, so that it can enter the validation phase and write phase safely. Deadlock can be avoided by enforcing each transaction to request its locks in the same sequence, e.g., based on the record key's order, so that no transaction waits for locks on new items while still locking other transactions' desired items.

`LogBase` delegates the task of managing distributed locks to a separate service, Zookeeper [2], which is widely used in distributed storage systems, such as Cassandra [16] and HBase [1], for providing efficient distributed synchronization. In addition, `LogBase` employs Zookeeper as a timestamp authority to establish a global counter for generating transaction's commit timestamps and therefore ensuring a global order for committed update transactions.

**Snapshot Isolation in LogBase.** The locking method during validation ensures "first-committer-wins" rule [4]. Therefore, the MVOCC in `LogBase` provides similar consistency and isolation level to standard snapshot isolation [4].

GUARANTEE 2. *Isolation. The hybrid scheme of multiversion optimistic concurrency control (MVOCC) in* `LogBase` *guarantees snapshot isolation.*

*Proof Sketch*: The MVOCC in LogBase is able to eliminate inconsistent reads, including "Dirty read", "Fuzzy read", "Read skew" and "Phantom", and inconsistent writes, including "Dirty write" and "Lost update", while still suffers from "Write skew" anomaly, thereby follows strictly the properties of Snapshot Isolation. Detailed proof could be found in [29]. □

The multiversion histories representing these phenomena when executing transactions in `LogBase` are listed below. In our notation, subscripts are used to denote different versions of a record, e.g., $x_i$ refers to a version of $x$ produced by transaction $T_i$. By convention, $T_0$ is an originator transaction which installs initial values of all records in the system.

Dirty read: $w_1[x_1]...r_2[x_0]...((c_1$ or $a_1)$ and $(c_2$ or $a_2)$ in any order)
Fuzzy read: $r_1[x_0]...w_2[x_2]...((c_1$ or $a_1)$ and $(c_2$ or $a_2)$– any order)
Read skew: $r_1[x_0]...w_2[x_2]...w_2[y_2]...c_2...r_1[y_0]...(c_1$ or $a_1)$
Phantom: $r_1[P]...w_2[y_2$ in $P]...c_2...r_1[P]...c_1$
Dirty write: $w_1[x_1]...w_2[x_2]...((c_1$ or $a_1)$ and $(c_2$ or $a_2)$ in any order)
Lost update: $r_1[x_0]...w_2[x_2]...w_1[x_1]...c_1$
Write skew: $r_1[x_0]...r_2[y_0]...w_1[y_1]...w_2[x_2]...(c_1$ and $c_2)$

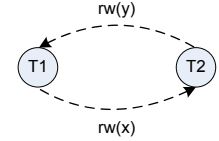

**Figure 5: Multiversion serialization graph for write skew.**

Under dependency theory [13], an edge from transaction $T_1$ to transaction $T_2$ is added into the multiversion serialization graph (MVSG) to represent their data conflicts in three scenarios: (1) $ww$-dependency where $T_1$ installs a version of $x$ and $T_2$ installs a later version of $x$, (2) $wr$-dependency where $T_1$ installs a version of $x$ and $T_2$ reads this (or a later) version of $x$, and (3) $rw$-dependency where $T_1$ reads a version of $x$ and $T_2$ installs a later version of $x$.

The MVSG of "Write skew", as depicted in Figure 5, contains a cycle between $T_1$ and $T_2$, showing that the MVOCC in `LogBase` suffers from this anomaly. On the contrary, the MVSG of the remaining phenomena (not shown) is acyclic, which means that `LogBase` is able to prevent those inconsistent reads and inconsistent writes. Therefore, `LogBase` provides snapshot isolation semantics for read-modify-write transactions.

Since snapshot isolation is a widely accepted correctness criterion and adopted by many database systems such as PostgreSQL, Oracle and SQL Server, we hypothesize that it is also useful for large-scale storages such as `LogBase`. If strict serializability is required, read locks also need to be acquired by transactions [27], but that will affect transaction performance as read locks block the writes and void the advantage of snapshot isolation. Another method which prevents cyclic "read-write" dependency at runtime is conservative and may abort transactions unnecessarily [6].

### 3.7.2 Commit Protocol and Atomicity

GUARANTEE 3. *Atomicity. The* `LogBase`*'s commit protocol guarantees similar atomicity property to the WAL+Data approach.*

The commit procedure for an update transaction $T$ proceeds as follows. After executing $T$'s read phase, the transaction manager runs the validation algorithm to determine if $T$ conflicts with other committed transactions or not. If the validation fails, then $T$ is restarted. Otherwise, the transaction manager gets a commit timestamp from the timestamp authority and persists $T$'s writes along with the commit record into the log repository. In addition, relevant in-memory index entries are updated accordingly to reflect the changes, and all the write locks held by $T$ are released.

Note that if the transaction manager fails to persist the final commit record into the log repository (due to errors of the log), $T$ is still not completed as in the WAL+Data approach. Although uncommitted writes could have been written to the log, they are not reflected in the index and thus cannot be accessed by users. Scan operations also check and only return data whose corresponding commit record exists. The uncommitted writes will be totally removed out of the log when the system performs log compaction. In summary, all or none of the updates of a transaction are recorded into the system, i.e., LogBase guarantees similar atomicity property to the WAL+Data approach.

Since the number of distributed transactions has been reduced at most by the use of smart data partitioning, the costly two-phase-commit protocol only happens in the worst case. LogBase further embeds an optimization technique that processes commit and log records in batches, instead of individual log writes, in order to reduce the log persistence cost and therefore improve write throughput. More details of the concurrency control and commit algorithm are presented in our technical report [29].

## 3.8 Failures and Recovery

We have shown how LogBase ensures atomicity, consistency and isolation property. In the following, we present the data durability property of LogBase, which guarantees all modifications that have been confirmed with users are persistent in the storage.

GUARANTEE 4. *Durability. The LogBase's recovery protocol guarantees similar data durability property to the WAL+Data approach.*

When a crash occurs, the recovery is simple in LogBase since it does not need to restore the data files as in the WAL+Data approach. Instead, the only instance in LogBase that needs to be recovered is the in-memory indexes. As a straightforward way, the restarted server can scan its entire log and rebuild the in-memory indexes accordingly. However, this approach is costly and infeasible in practice. In order to reduce the cost of recovery, LogBase performs checkpoint operation at regular times or when the number of updates has reached a threshold.

In the checkpoint operation, tablet servers persist two important information into the underlying DFS to enable fast recovery. First, the current in-memory indexes are flushed into index files stored in DFS for persistence. Second, necessary information, including the current position in the log and the log sequence number (LSN) of the latest write operation whose effects have been recorded in the indexes and their persisted files in the first step, are written into checkpoint blocks in DFS so that LogBase can use this position as a consistent starting point for recovery.

With the checkpoint information, recovery from machine failures in LogBase can be performed fast since it only needs to do an analysis pass from the last known consistent checkpoint towards the end of the log where the failures occurred. At restart time the tablet server can reload the indexes quickly from the persisted index files back into the memory. Then a redo strategy is employed to bring the indexes up-to-date, i.e., the tablet server analyzes the log entries from the recovery starting point and updates the in-memory indexes accordingly. If the LSN of the log entry is greater than the corresponding index entry in the index, then the pointer in the index entry is updated to this log address. Performing redo is sufficient for system recovery since LogBase adopts optimistic concurrency control method, which defers all modifications until commit time. All uncommitted log entries are ignored during the redo process and will be discarded when the system performs log compaction. In addition, in the event of repeated restart when a crash occurs during the recovery, the system only needs to redo the process.

Note that if a tablet server fails to restart within a predefined period after its crash, the master node will consider this as permanent failures and re-assign the tablets maintained by this failed server to other healthy tablet servers in the system. The log of the failed servers, which is stored in the shared DFS, is scanned (from the consistent recovery starting point) and split into separate files for each tablet according to the tablet information in the log entries. Then the healthy tablet servers scan these additional assigned log files to perform the recovery process as discussed above.

## 4. PERFORMANCE EVALUATION

### 4.1 Experimental Setup

Experiments were performed on an in-house cluster including 24 machines, each with a quad core processor, 8 GB of physical memory, 500 GB of disk capacity and 1 gigabit ethernet. LogBase is implemented in Java, inherits basic infrastructures from HBase open source, and adds new features for log-structured storages including access to log files, in-memory indexes, log compaction, transaction management and system recovery. We compare the performance of LogBase with HBase (version 0.90.3). All settings of HBase are kept as its default configuration, and LogBase is configured to similar settings. Particularly, both systems use 40% of 4 GB heap memory for maintaining in-memory data structures (the memtables in HBase and in-memory indexes in LogBase), and 20% of heap memory for caching data blocks. Both systems run on top of Hadoop platform (version 0.20.2) and store data into HDFS. We keep all settings of HDFS as default, specifically the chunk size is set to 64 MB and the replication factor is set to 3.

Each machine runs both a data node and a tablet server process. The size of datasets is proportional to the system size, and for every experiment we bulkload 1 million of 1KB records for each node (the key of each record takes its value from $2 * 10^9$ which is the max key in YCSB benchmark [10]). For scalability experiments, we run multiple instances of benchmark clients, one for each node in the system. Each benchmark client submits a constant workload into the system, i.e., a completed operation will be immediately followed by a new operation. The benchmark client reports the system throughput and response time after finishing a workload of 5,000 operations. Before running every experiments, we execute about 15,000 operations on each node to warm up the cache. The default distribution for the selection of accessed keys follows Zipfian distribution with the co-efficient set to 1.0.

### 4.2 Micro-benchmarks

In this part, we study the performance of basic data operations including sequential write, random read, sequential scan and range scan of LogBase with a single tablet server storing data on a 3-node HDFS. We shall study the performance of LogBase with mixed workloads and bigger system sizes in the next section.

*4.2.1 Write Performance*

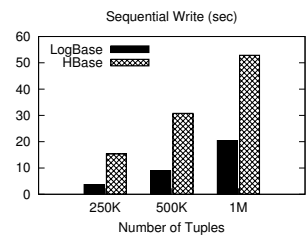

**Figure 6: Write performance.**

Figure 6 plots the write overhead of inserting 1 million records into the system. The results show that LogBase outperforms HBase by 50%. For each insert operation, LogBase flushes it to the log and then update the memory index. It thus only writes the data to HDFS once. On the contrary, besides persisting the log information (which includes the record itself) into HDFS, HBase has to insert the record into a memtable, which will be written to the data file in HDFS when the memtable is full (64 MB as default setting). As a result, HBase incurs more write overhead than LogBase.

### 4.2.2 Random Access Performance

Figure 7 shows the performance of random access without any cache used in both systems. The performance of LogBase is superior to HBase, because LogBase maintains a dense in-memory index and each record has a corresponding index entry containing its location in the log. With this information, LogBase is able to seek directly to the appropriate position in the log and retrieve the record. In contrast, HBase stores separate sparse block indexes in different data files, and hence after seeking to the corresponding block in one data file, it loads that block into memory and scan the block to get the record of interest. Further, the tablet server in HBase has to check its multiple data files in order to get the proper data record. Therefore, LogBase can efficiently support long tail requests that access data not available in the cache.

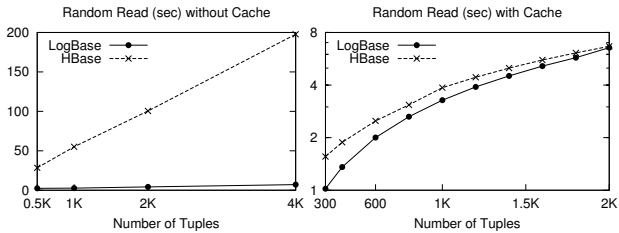

**Figure 7: Random access (without cache).** **Figure 8: Random access (with cache).**

As shown in Figure 8, the performance gap between LogBase and HBase reduces when the block cache is adopted in the system. The main reason is that, if the block containing the record to be accessed is cached from previous requests, HBase does not need to seek and read the entire block from HDFS. Instead, it only reads the proper record from the cached block. Note that with larger data domain size in distributed YCSB benchmark as will be discussed in the next section, the cache has less effect and LogBase provides better read latency for the support of in-memory indexes.

### 4.2.3 Scan Performance

**Sequential scan.** Figure 9 illustrates the result of sequential scan the entire data. The performance of LogBase is slightly slower than HBase. LogBase scans the log files instead of the data files as HBase, and each log entry contains additional log information besides the data record such as the table name and column group. As such, a log file has larger size than a data file and LogBase has to spend slightly more time to scan the log file.

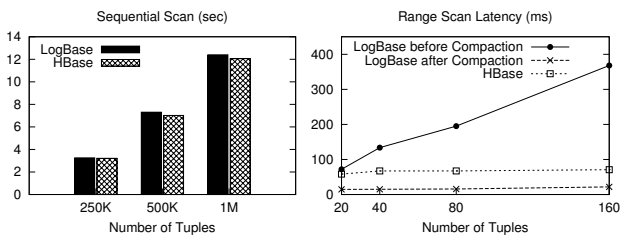

**Figure 9: Sequential scan.** **Figure 10: Range scan.**

**Range scan.** The downside of LogBase is that it is not as efficient as HBase when processing range scan query as shown in Figure 10. In HBase, data in memtables are kept sorted by key order and persisted into data files, and hence facilitates fast range scan query. LogBase, on the contrary, sequentially writes data into the log without any clustering property and might need to perform multiple random access to process a single range scan query. However, it is notable that after the compaction process, data in the log are well-clustered and LogBase is able to provide even better range scan performance than HBase for its ability to load the correct block quickly with the support of dense in-memory indexes.

## 4.3 YCSB Benchmark

In the following, we examine the efficiency and scalability of LogBase with mixed workloads and varying system sizes using YCSB benchmark [10]. The system size scales from 3 to 24 nodes and two write-heavy mix workloads (95% and 75% of update in the workload) are tested.

In the loading phase of the benchmark, multiple instances of clients are launched to insert benchmark data in parallel. Similar to the result of sequential write in the micro-benchmark, Figure 11 shows that LogBase outperforms HBase when parallel loading data and only spends about half of the time to insert data. This confirms that LogBase can provide highly sustained throughput for write-heavy environments.

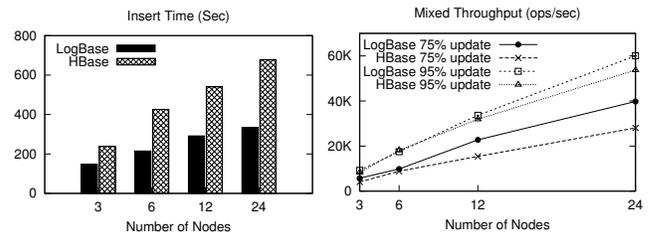

**Figure 11: Data loading time.** **Figure 12: Mixed throughput.**

In the experiment phase, the benchmark client at each node will continuously submit a mixed workload into the system. An operation in this workload either reads or updates a certain record that has been inserted in the loading phase. The system overall throughput with different mixes is plotted in Figure 12 and the corresponding latency of update and read operations is shown in Figure 13 and Figure 14 respectively. The results show that both LogBase and HBase achieve higher throughput with the mix that has higher percentage of update since both systems perform write operations more efficient than read operations.

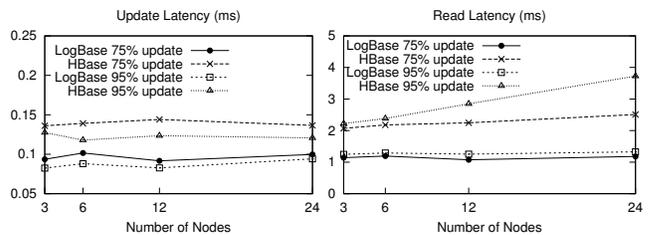

**Figure 13: Update latency.** **Figure 14: Read latency.**

In addition, for each mix, LogBase achieves higher throughput than HBase for its ability to support both write and read efficiently. In HBase, if the memtable is full and a minor compaction is required, the write has to wait until the memtable is persisted successfully into HDFS before returning to users and hence the write response time is delayed. LogBase provides better read latency for the support of in-memory indexes as we have shown in the micro-benchmarks. Although HBase employs cache to improve read performance, the cache has less effect in this distributed experiment since both data domain size and experimental data size are large, which affects read performance.

Figure 13 and Figure 14 also illustrate the elastic scaling property of LogBase where the system scales well with flat latency. That is, the more workload can be served by adding more nodes into the system.

## 4.4 TPC-W Benchmark

In this experiment, we examine the performance of `LogBase` when accessing multiple data records possibly from different tables within the transaction boundary. In particular, we experiment `LogBase` with TPC-W benchmark which models a webshop application workload. The benchmark characterizes three typical mixes including browsing mix, shopping mix and ordering mix that have 5%, 20% and 50% update transactions respectively.

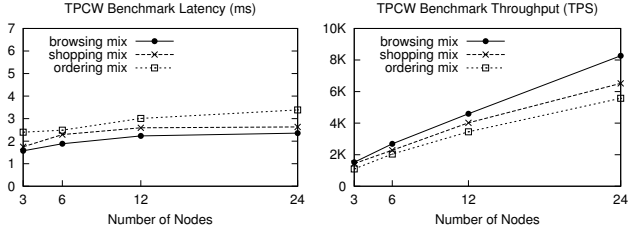

**Figure 15: Transaction latency.**  **Figure 16: Transaction throughput.**

A read-only transaction performs one read operation to query the details of a product in the `item` table while an update transaction executes an order request which bundles one read operation to retrieve the user's `shopping cart` and one write operation into the `orders` table. Each node in the system is bulk loaded with 1 million products and customers before the experiment. We stress test the system by using a client thread at each node to continuously submit transactions to the system and then benchmark the transaction throughput and latency.

As can be seen in Figure 15, under browsing mix and shopping mix, `LogBase` scales well with nearly flat transaction latency when the system size increases and as a result, the transaction throughput (shown in Figure 16) scales linearly under these two workloads. The low overhead of transaction commit is attributed to this result since in these two workloads, most of the transactions are read-only and always commit successfully without the need of checking conflicts with other transactions for the use of MVOCC.

## 4.5 Checkpoint and Recovery

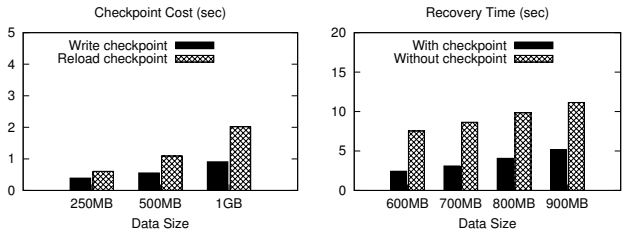

**Figure 17: Checkpoint cost.**  **Figure 18: Recovery time.**

We now study the cost of checkpoint operation and the recovery time in a system of 3 nodes. Figure 17 plots the time to write a checkpoint and reload a checkpoint with varying thresholds at which a tablet server performs the checkpoint operation. `LogBase` takes less time to write a checkpoint (persist in-memory indexes) than to reload a checkpoint (reload the persisted index files into memory) because HDFS is optimized for high write throughput. This is useful because checkpoint writing is to be performed more frequently in `LogBase`, whereas checkpoint loading only happens when the system recovers from tablet servers' failures.

The time to recover varying amount of data maintained by a failed tablet server is shown in Figure 18. The checkpoint was taken at a threshold of 500 MB before we purposely killed the tablet server when its amount of data reached 600 MB to 900 MB. The results show that the recovery time in the system with checkpoint is significantly faster than without checkpoint. In the former approach, the system only needs to reload the checkpoint and scan a little additional log segments after the checkpoint time to rebuild the in-memory indexes, whereas in the latter approach the system has to scan the entire log segments.

`LogBase` does not support as efficient recovery time as RAMCloud [22] because the two systems make different design choices for targeting at different environments. In RAMCloud, both indexes and data are entirely stored in memory while disks only serve as data backup for recovery purpose. Therefore, RAMCloud backups log segments of a tablet dispersedly to hundreds of machines (and disks) in order to exploit parallelism for recovery. In contrast, `LogBase` stores data on disks and hence cannot scatter log segments of a tablet to such scale in order to favor recovery as it would adversely affect the write and read performance of the system.

## 4.6 Comparison with Log-structured Systems

As we have reviewed in Section 2, recent scalable log-structured record-oriented systems (LRS) such as RAMCloud [22] and Hyder [5] target at different environments with `LogBase`. Specifically, RAMCloud stores its data and indexes entirely in memory while Hyder scales its database in shared-flash environments without data partitioning. Therefore, we cannot compare their performance directly with `LogBase`. Here, for comparison purpose as well as exploring the opportunity of scaling the indexes beyond memory, we examine a system, referred to as LRS, which has a distributed architecture and data partitioning strategy similar to RAMCloud and `LogBase` but stores data on disks and indexes them with log-structured merge trees (LSM-tree) [21] to deal with scenarios where the memory of tablet servers is scarce. Particularly, in this experiment we use LevelDB [4], a variant LSM-tree open source by Google, with all settings kept as default.

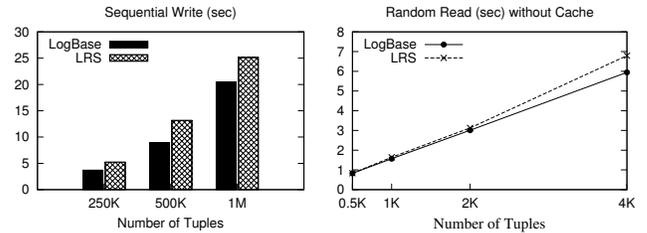

**Figure 19: Sequential write.**  **Figure 20: Random access.**

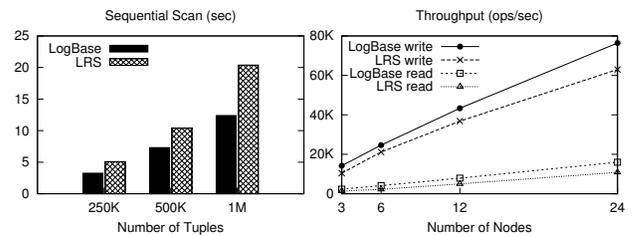

**Figure 21: Sequential scan.**  **Figure 22: Throughput.**

The results of comparison between `LogBase` and LRS in a system of 3 nodes are shown in Figure 19, Figure 20, and Figure 21 respectively for sequential write, random access, and sequential scan. The comparison results with varying system sizes are also plotted in Figure 22. Overall, the sequential write and random access performance of LRS are only slightly lower than that of `LogBase` because LevelDB is highly optimized for a variety of workloads and can provide efficient write and read performance with moderate write and read buffer (4 MB and 8 MB respectively in the experiment). This leads us to conclude that it is possible for `LogBase` to scale its indexes beyond memory (by the use of LSM-trees) without paying much cost of reduction in the system throughput.

---
[4] http://code.google.com/p/leveldb/

`LogBase` also achieves higher sequential scan performance than LRS. Recall that for each scanned record, the system needs to check its stored version against the current version maintained in the indexes to determine whether the record contains the latest data. Such cost of accessing indexes is attributed to the difference in the scan performance of the two systems. Note that after log compaction, historical versions of a record are clustered together and hence the number of version checking with indexes is minimized, which would reduce the scan performance gap.

## 5. CONCLUSION

We have introduced a scalable log-structured database system called `LogBase`, which can be elastically deployed in the cloud and provide sustained write throughput and effective recovery time in the system. The in-memory indexes in `LogBase` support efficient data retrieval from the log and are especially useful for handling long tail requests. `LogBase` provides the widely accepted snapshot isolation for bundled read-modify-write transactions. Extensive experiments on an in-house cluster verifies the efficiency and scalability of the system. Our future works include the design and implementation of efficient secondary indexes and query processing for `LogBase`.

## Acknowledgments

This work was in part supported by the Singapore MOE Grant No. R252-000-454-112. We would like to thank anonymous reviewers and Hank Korth for their insightful feedback, and Yuting Lin for initial system design and implementation support.